\documentstyle[12pt]{article}
\newcommand {\qed}{$\Box$}
\newcommand{\U}{\:\:{\sf Until}\:\:}
\newcommand{\X} {\:\:{\sf Nexttime}\:\:}
\begin{document}
\title{Formal Languages and Algorithms for Similarity based Retrieval from 
Sequence Databases
\thanks{This research is supported in part by the NSF grants CCR-0205365, CCR-9988884
and IIR 9711925. A preliminary version of this paper appeared as \cite{Si02}.} 
}

\author{ A. Prasad Sistla\\
Department of Computer Science,\\
University of Illinois at Chicago,\\
Chicago, Illinois 60680,\\
sistla@cs.uic.edu
       }
\maketitle
\begin{abstract}
The paper considers various formalisms based on Automata, Temporal Logic
and Regular expressions for specifying queries  over finite sequences. 
Unlike traditional semantics that associate  $true$ or $false$ value denoting
whether a sequence satisfies a query, the paper presents  distance 
measures that associate a value in the interval ${\tt [0,1]}$ 
with a sequence and a query, denoting how closely the sequence satisfies the 
query. These measures are defined using a spectrum of normed vector distance
measures. Such similarity based semantics can be used for retrieval of database
sequences that approximately satisfy a query.
Various measures based on the syntax of the query and the traditional
semantics of the query are presented. Efficient Algorithms for computing 
these distance
measures are presented.
\end{abstract}  
\section{Introduction}
Recently there has been much interest in  similarity based retrieval 
from sequence databases. The works in this area, consider a database
of sequences and  provide methods for retrieving sequences that approximately
match a given query.
 Such methods can be applied for retrieval from
time-series, video and textual databases.
Earlier work in this area \cite{FRM94} considered the case when the query is
given by a single sequence and developed fast methods for retrieving 
all database sequence that closely match the 
query sequence. In this paper, we consider the case when the query is
given by a predicate over sequences specified in various formalisms,
 and give efficient methods for
checking if a database sequence approximately satisfies the query predicate.

We consider formalisms based on automata,  temporal logic,
and regular expressions for specifying queries over sequences.
We define similarity based semantics for these formalisms.
More specifically, for a database sequence $d$ and  query $q$,
we define a similarity
measure  that denotes how closely $d$ satisfies the
query $q$. This measure ranges from zero to one denoting the different levels
 of satisfaction; higher values denote greater levels of satisfaction with
 value one indicating perfect satisfaction. Actually, we define
a distance measure between $d$ and $q$,
and define the similarity measure to be (1 - distance measure).
(Both the distance and the similarity measures normally take values in the 
interval {\tt [0, 1]}. However, if there is no sequence of length $d$
that satisfies $q$ then the distance meaure can take the value $\infty$
and the similarity value is $-\infty$ in this case).

For example, in a database consisting of daily stock sequences, one might
request a query such as--- ``retrieve all the daily stock patterns in which
IBM stock price remained below 70 until the Dow-Jones value reached 10,000''.
 Such a query can be expressed in Temporal Logic (or any of the other 
formalisms) by considering 
``IBM stock price is less than 70'' and ``Dow-Jones value equals 10,000''
as atomic propositions. An answer to such a query will not only return 
sequences that exactly satisfy the temporal predicate, but also those
sequences that satisfy it approximately, i.e., those sequences having
a similarity value greater than a given threshold specified by the user.

The distance measures that we define are classified in to 
 semantics based and syntax based measures. 
We define two types of semantic based distance measures.
In both of these types,
the distance measure is defined using the exact semantics of the query,
which is given by a set $S$ of sequences.
When the query $q$ is given by an automaton then $S$ is the set of sequences
accepted by the automaton; when $q$ is a temporal logic formula, then $S$ is the set
of (finite) sequences that satisfy the formula; when $q$ is a regular
expression, then $S$ is the language specified by the regular expression.
The first semantics based distance measure between $d$ and $q$,
is defined to be the minimum of the vector distances between $d$ and
each sequence in the set $S$. 
The second distance semantic measure is more complex and is based on replacing 
certain symbols in sequences of $S$ by the wild card symbols (section \ref{def-sec}
contains the actual definition). 
By using various  norm vector distance functions, we get a spectrum of the two
types of semantic distance measures.

The syntax based distance measure is defined only for the cases when the query is specified
by a temporal formula or by a regular expression.
In this case,the distance measure
is defined inductively based on the syntax of the query, i.e.
it's value is defined as a function of the distance
measures of $d$ with respect to the top level components of $q$ (i.e.,
sub-formulas when $q$ is a temporal logic formula). For example,
the syntax distance with respect to the temporal formulas
$g \wedge h$ is defined to be the maximum of the distances 
with respect to $g$ and $h$.
We relate the syntax and semantics based distance measures. 

We present algorithms for computing the syntactic and semantic distances 
for a given database sequence and a query. 
For the case when the query is given by an automaton or by a regular
expression, the algorithms for 
the first semantic distance measure have linear complexity  in the size of the
automaton  and polynomial complexity in the length of the sequence (actually, 
the complexity is linear in the length of the sequence for the infinite norm,
and is quadratic for other cases); for the second semantic distance measure the
algorithms have the same complexity with respect to the length of the sequence,
but have triple exponential complexity in terms of the automaton size.
When the query is given by a temporal logic formula, the algorithms  for
the semantic distance measures have the same complexity with the following exception:
the first semantic distance measure has exponential complexity in the
length of the formula; this blow up is caused by the translation of the temporal
formula in to an automaton.

The algorithms for computing syntactic distance measures have
linear complexity in the length of the query and polynomial
complexity in the length of the database sequence;
(more specifically, the complexity with respect to the database sequence is
linear for the infinite norm vector distance function, and is quadratic
in other cases).

The paper is organized as follows. Section \ref{def-sec} gives definitions 
and notation.
Section \ref{aut-sec} reviews and presents algorithms for the case when the
query is specified by automata.
Section \ref{temp-sec} defines the distance measures for temporal logic 
and presents
algorithms for computing these. Section \ref{reg-sec} presents the 
corresponding results
for the regular expressions. Section \ref{rel-sec} briefly discusses 
related work. Section \ref{conc-sec} contains conclusions. 

\section{Definition and Notation}
\label{def-sec}

In this section, we define the various formalisms that we consider in the paper
and their similarity based semantics.
For a sequence $s\:=\:(s_0,s_1,...,s_i,...s_{n-1})$, we let ${\tt s[i]}$
denote its suffix starting from $s_{i}$. 
We let $length(s)$ denote the length of $s$. A null sequence is a sequence of length zero.
If $s,t$ are sequences, then
we let $st$ denote the concatenation of $s$ and $t$ in that order.
We represent a sequence having only one element by that element.
A sequence over a set $\Delta$ is a sequence whose elements are from $\Delta$.
A language $L$ over $\Delta$ is a set of sequences over $\Delta$. 
We let $\Delta^*$ represent the set of all such sequences.
If $L,M$ are languages over $\Delta$ then $LM$ and $L^*$ are languages
over $\Delta$ defined as follows: 
$LM\:=\{\alpha\beta\::\alpha\in L, \beta\in M \}$; 
$L^*\:= \cup_{i\geq 0} L^i$; here $L^i$ is the concatenation of $L$ with itself
$i$ times; $L^0$ is the singleton set containing the empty string.
For a language $L$ over $\Delta$, we let $\overline{L}$ be the complement of
$L$, i.e., $\overline{L}\:=\Delta^* - L$.
Whenever there is no confusion, we represent a set containing a single element by that
element itself.
 
Let $\Delta$ be a set of elements, called atomic queries.
Each member of $\Delta$ represents an atomic query on a database state. 
With each database state $u$ and atomic query $a$, we associate
a similarity value $simval(u,a)$ that denotes how closely
$d$ satisfies $a$. This value can be any value between zero and one (one
indicates perfect satisfaction).
We use a special atomic query $\phi\notin \Delta$, called wild card, which
is always satisfied in every database state; that is, $simval(u,\phi)=1$
for every database state $u$.
Let $d\:=\:(d_0,d_1,...,d_{n-1})$ be a sequence of database states
and  $a\:=\:(a_0,...,a_{n-1})$ be a sequence over
$\Delta \cup \{\phi\}$. Corresponding to $d$ and $a$, 
we define a sequence of real numbers called 
$simvec(d,a)$ defined as follows. Let $i_0<i_1<...<i_{m-1}$
be all values of $j$ such that $0\leq j<n$ and $a_j\neq \phi$.
We define $simvec(d,a)$ to be the sequence
$(x_0,...,x_{m-1})$
where $x_j\:=\:simval(d_{i_{j}},a_{i_{j}})$ for $0\leq j<m$.
Intuitively, we define $simvec(d,a)$ by ignoring the positions corresponding
to the wild card symbol.
It is to be noted that if $a$ contains only $\phi$ then
$simvec(d,a)$ is the empty sequence, i.e. it is of length zero.

Let $F$ be a distance measure over real vectors assigning a positive
real value less than or equal to one,
i.e., $F$ is a function which associates a real value 
$F(\vec{x},\vec{y})$, such that $0\leq F(\vec{x},\vec{y})\leq 1$,
 with every pair of real vectors  $\vec{x},\vec{y}$ of same length.
Given a database sequence $d$ and a sequence $a$ of equal length over $\Delta\cup\{\phi\}$,
we define $dist(d,a,F)$ as follows: if $d$ and $a$ are of different lengths
then $dist(d,a,F)\:=\infty$; if $d$ and $a$ are of the same length and
$simvec(d,a)$
is not the empty sequence then $dist(d,a,F)\:=F(simvec(d,a),\vec{1})$
where 
 $\vec{1}$ denotes a vector, of the same length as $simvec(d,a)$,
all of whose components are 1; if $d$ and $a$ are of the same length
and $simvec(d,a)$ is the empty sequence
then $dist(d,a,F)$ is defined to be $0$. 
Thus it is to be seen that
if $d$ and $a$ are of the same length then $dist(d,a,F)$ has value
between $0$ and $1$, otherwise it has value $\infty$. 
The value of $\infty$ is given, in the later case, in order to distinguish it
from the case when $d$ and $a$ have equal lengths and 
$F(simvec(d,a),\vec{1})=1$ and also for technical convenience.

Let $L$ be a language over $\Delta$, $d$ be a database sequence, and
$F$ be a vector distance function.
We define two distance measures, $distance_1(d,L,F)$ and $distance_2(d,L,F)$,
 of $d$ with respect to $L$  using the
vector distance function $F$. If $L$ is non-empty then
$distance_1(d,L,F)$ is defined to be 
$\min \{dist(d,a,F):\:a\in L\:$ and $length(a)=length(d)\}$, 
otherwise it is defined to be $\infty$. It is to be noted that
if $L$ does not contain any strings of the same length as $d$ then
$distance_1(d,L,F)$ is $\infty$. 

The definition of $distance_2(d,L,F)$ is more
complex and is motivated by the following situation. Suppose each atomic 
query in
$\Delta$ denotes a logical predicate and the disjunction of all these
predicates is a tautology, i.e. it is always satisfied. As an example,
consider the case when $\Delta\:=\{P, \neg P\}$ where $P$ is an atomic 
proposition. 
Now consider the language
$L_1\:=\:\{aP:\:a\in \Delta\}$.
The language $L_1$ requires that the atomic proposition  $P$ be satisfied 
at the second state irrespective of whether $P$ is satisfied or not at the
first state. It can be argued that the distance
of a database sequence $d$ with respect to $L_1$ should depend only
on the distance of the second database state with respect to $P$.
This intuition leads us to the following definitions.

Recall that $L$ is a language over $\Delta$.
Let $closure(L)$ be the smallest language $L'$ over 
$\Delta \cup \{\phi\}$ such that $L\subseteq L'$ and  the 
following closure condition is satisfied for every 
$\alpha,\beta\in (\Delta\cup\{\phi\})^*$: if $\alpha a \beta\:\in L'$
for every $a\in \Delta$ then $\alpha\phi\beta\:\in L'$. (Recall $\phi$
is the wild card symbol).
Now we define a partial order $<$ on the set of sequences over 
$\Delta\cup\{\phi\}$ as follows.
Let $\alpha\:=\:(\alpha_0,...,\alpha_{m-1})$ and
$\beta\:=\:(\beta_0,...,\beta_{n-1})$ be any two sequences 
over $\Delta\cup\{\phi\}$.
Intuitively, $\alpha < \beta$ if $\alpha$ can be obtained from
$\beta$ by replacing some of the occurrences of the wild card symbol 
$\phi$ in $\beta$ by a symbol in $\Delta$. 
Formally,  $\alpha< \beta$ iff $m=n$ and for each $i=0,1,...,n-1$ 
either $\alpha_i= \beta_i$ or $\beta_i=\phi$, and 
there exists at least one value
of $j$ such that $\alpha_j\in \Delta$ and $\beta_j\:=\phi$.

 It is not difficult to see  that
$<$ is a partial order. Let $S$ be any set of sequences over
$\Delta\cup\{\phi\}$. A sequence $\alpha\in S$ is called maximal
if there does not exist any other sequence $\beta\in S$ such that
$\alpha<\beta$. Let $maximal(S)$ denote the set of all maximal 
sequences in $S$.
Now we define $distance_2(d,L,F)$ to be the
value of $distance_1(d,\:maximal(closure(L)),F)$.
For the language $L_1$ (given above) $maximal(closure(L))$ consists
of the single sequence $\phi P$;
for a database sequence $d$ of length two, 
it should be easy to see that $distance_2(d,L_1,F)$ equals the distance of
the second database state with respect to $P$. 

It is to be noted that the distance functions $distance_1$ and $distance_2$
depend on the vector distance function $F$. 
We consider a spectrum of vector distance functions 
$\{F_k\::k=1,2,...,\infty\}$
defined as follows. Let $\vec{x},\vec{y}$ be two vectors of length $n$.
The value of $F_k(\vec{x},\vec{y})$
is given as follows.
For $k\neq \infty$,
\begin{equation}F_k(\vec{x},\vec{y})\:=\:(\frac{\Sigma_{0\leq i< n}(|x_i-y_i|)^k}{n})^{1/k} \label{eq1} \end{equation}
\begin{equation}F_\infty(\vec{x},\vec{y})\:=\:\max \{|x_i-y_i|\::\:0\leq i< n\} \label{eq2} \end{equation}
Note that $F_1$ is the average block distance function
and $F_2$ is the mean square distance function, etc.
It can easily be shown that $F_\infty(\vec{x},\vec{y})\:=\:\lim_{k\rightarrow \infty} F_k(\vec{x},\vec{y})$.
Note that $F_1(\vec{x},\vec{y})$ gives equal importance to all
components of the vectors; however, as $k$ increases,  the numerator
in the expression for $F_k(\vec{x},\vec{y})$ is dominated by
the term having the maximum value, i.e.  
by $\max \{|x_i-y_i|\::\:1\leq i\leq n\}$, and in the limit 
$F_\infty(\vec{x},\vec{y})$ equals this maximum value.
Thus we see that $F_1$ and $F_\infty$ are the extremes of our distance 
functions. We call $F_1$ as the {\it average block distance} function
and $F_\infty$ as the {\it infinite norm} distance function.

The following lemma can easily be proved from our earlier observations.

LEMMA 2.1: For every database sequence $d$ and language $L$ over $\Delta$
the following properties hold.
\begin{enumerate}
\item $distance_{2}(d,L,F_{\infty})\leq\: distance_{1}(d,L,F_{\infty})$.
\item For any $i,j\in \{1,2,...,\infty\}$ such that $i<j$, 
$distance_{1}(d,L,F_i)\leq\: distance_{1}(d,L,F_j)$ and 
$distance_{2}(d,L,F_i)\leq\: distance_{2}(d,L,F_j)$.
\end{enumerate} 

{\bf Proof:} Part 1 of the lemma is proved by the following argument.
Let $length(d)=n$.
If $\alpha,\beta$ are two sequences of length $n$ over $\Delta \cup \{\phi\}$
such that $\alpha < \beta$ or $\alpha =\beta$ then 
$dist(d,\alpha,F_\infty) \geq dist(d,\beta,F_\infty)$ (To see this, let
$\alpha = (\alpha_0,...,\alpha_i,...,\alpha_{n-1})$ and 
$\beta =  (\beta_0,...,\beta_i,...,\beta_{n-1})$; since, for each 
$i$, $0\leq i<n$, either $\alpha_i=\beta_i$ or $\beta_i = \phi$, and
$simval(d_i,\phi)=1$,
it is the case that $(1 -simval(d_i,\alpha_i))\geq (1-simval(d_i,\beta_i))$;
Since $dist(d,\alpha,F_\infty) = \max \{(1-simval(d_i,\alpha_i)) :0\leq i<n\}$
and $dist(d,\beta,F_\infty) = \max \{(1-simval(d_i,\beta_i)) :0\leq i<n\}$,
it follows that $dist(d,\alpha,F_\infty)\geq dist(d,\beta,F_\infty)$;
it is to be noted that this relation will not hold, in general, if we
replace $F_\infty$ by $F_k$ for any $k<\infty$).
From the definition of $maximal(closure(L))$, we see that, for every 
$\alpha\in L$ there exists a string $\beta\in maximal(closure(L))$ such that
$\alpha <\beta$ or $\alpha =\beta$. From this we see that, for every $\alpha\in L$ 
of length $n$, there exists a string $\beta\in maximal(closure(L))$ of length $n$
such that $dist(d,\alpha,F_\infty)\geq  dist(d,\beta,F_\infty)$. As a consequence,
$\min \{dist(d,\alpha,F_\infty)\:: \alpha \in L\}$ 
$\geq \min \{ dist(d,\beta,F_\infty)\:: \beta\in maximal(closure(L))\}$.
Now from the definitions, we see that the the left and right hand sides of the
above inequality
are exactly $distance_1(d,L,F_\infty)$ and $distance_2(d,L,F_\infty)$
respectively. Part 1 of the lemma follows from this.
Part 2 of the lemma follows from the well known fact that 
for any two $n$-vectors vectors $\vec{a},\vec{b}$, each of whose components lie in the
positive unit interval, $F_i(\vec{a},\vec{b})\leq F_j(\vec{a},\vec{b})$ for $i<j$. 
$\Box$

It is to be noted that part 1 of the lemma does not hold, in general,
if we replace $F_\infty$ by $F_k$ for any $k<\infty$.
The following is a simple counter example for this. Let $\Delta\:=\{a,b\}$
and  $L\:=\{ab, bb\}$. Let $d=(d_0,d_1)$. Assume that
$simval(d_0,a)$ and $simval(d_0,b)$ be both equal to $\frac{1}{2}$,
and $simval(d_1,b)=0$. It should be easy to see that 
$distance_1(d,L,F_1)= \frac{3}{4}$. It should also be easy to see that
$maximal(closure(L))$ contains the single string $\phi b$. Hence
$distance_2(d,L,F_1)\:=dist(d,\phi b,F_1)$ which equals $1$.
Thus, in this case, part 1 of the lemma does not hold when we use
$F_1$. In fact, in this case, part 1 of the lemma does not hold for any
$F_k$ where $k<\infty$. On the contrary,
$distance_1(d,L,F_1)\leq distance_2(d,L,F_1)$.
We can also give an example for which $distance_2(d,L,F_1)\leq distance_1(d,L,F_1)$.
Thus, in general, $distance_1(d,L,F_k)$ and $distance_2(d,L,F_k)$ are not related for any
$k<\infty$.

\section{Automata}
\label{aut-sec}
In this section, we consider automata for specifying queries over sequences.
We give algorithms for computing the two distances of a database sequence 
with respect to a given automaton.

An automaton ${\cal A}$ is 5-tuple $(Q,\Delta,\delta,I,Final)$ where $Q$
is a finite set of states, $\Delta$ is a finite set of symbols
called the input alphabet, $\delta$ is the set of transitions,
$I,Final \subseteq Q$ are the set of initial and final states, respectively.
Each transition of ${\cal A}$, i.e. each member of $\delta$, is a triple
of the form $(q,a,q')$ where $q,q'\in Q$ and $a\in \Delta$; this triple
denotes that the automaton makes a transition from state $q$ to $q'$
on input $a$; we also represent such a transition as $q\;\rightarrow_a\;q'$.
Each input symbol represents an atomic predicate (also called
an atomic query in some places) on a single database state.
For example, in a stock market database, $price(ibm)=100$ represents
an atomic predicate. In a textual database, each database state
represents a document and a database sequence represents a sequence
of documents; here an atomic predicate
may state that the document contain some given key words.
In a video database, which is a sequence of images or shots, each
atomic predicate represents a condition on a picture such as
requiring that the picture contain some given objects.

Let $a\:=\:a_0,a_1,...,a_{n-1}$ be a sequence of input symbols from
$\Delta$ and $q,q'$ be states in $Q$. We say that the sequence $a$
takes the automaton ${\cal A}$ from state $q$ to $q'$ if there exists
a sequence of states $q_0,q_1,...,q_n$  such that $q_0=q$ and $q_n=q'$
and for each $i=0,...,n-1$, $q_i\:\rightarrow_{a_{i}}q_{i+1}$ 
is a transition of ${\cal A}$.
For any state $q$, we let $T(q)$ denote the set of sequences that take
the automaton from state $q$ to a final state.
We say that the automaton ${\cal A}$ accepts the string $a$ if there exists
an initial state $q$ such that $a\in T(q)$.
We let $L({\cal A})$ denote the set of strings accepted by ${\cal A}$.
We let $|{\cal A}|$ denote the number of its states,  i.e., that cardinality
of $Q$, and $Size({\cal A})$ denote the sum of its number of states and transitions,
i.e., the sum of the cardinalities of $Q$ and $\delta$. 

We identify vectors with sequences. We let $\vec{1}$ denote
a vector all of whose components are 1. The length of such a vector will
be clear from the context.

For a database sequence $d$, automaton ${\cal A}$, and a vector distance
function $F$, we define two  distances $distance_i(d,{\cal A},F)$,
for $i=1,2$ as follows: $distance_i(d,{\cal A},F)$ 
$\:= distance_i(d, L({\cal A}),F)$.
\\
\noindent{\bf Algorithms for Computing the distances}

Now we outline an algorithm for computing the value of $distance_1(d,{\cal A}, F_{\infty})$.
Recall that $F_{\infty}$ is the infinity norm vector distance measure.
Let $d=(d_0,...,d_{n-1})$ be the database sequence.
Essentially, the algorithm computes the distances of the suffixes of $d$ with respect
to the states of the automaton for
increasing lengths of the suffixes.  
For any automaton state $q$ and integer $i$ ($0\leq i\leq n-1$),
the algorithm first computes the value of $distance_1({\tt d[i]},T(q),F_{\infty})$ 
in decreasing values of $i$ starting with $i=n-1$. (Recall that ${\tt d[i]}$ is the
suffix of $d$ starting with $d_i$ and $T(q)$ is the set of sequences accepted by ${\cal A}$
starting in the state $q$). The algorithm finally computes
$distance_1(d,{\cal A},F_{\infty})$ to be minimum of the values in
$\{ distance({\tt d[0]},T(q),F_{\infty})\:: q\in I\}$; note that ${\tt d[0]}$ is
simply $d$. 
The values in the set $\{distance_1({\tt d[i]}, T(q), F_{\infty})\::q\in Q\}$
are computed in decreasing values of $i$ using the recurrence equation given by the
following lemma.

LEMMA 3.1: Let $q$ be any state in $Q$ and
 $q\rightarrow_{a_{1}}q_1,...,$$q\rightarrow_{a_{m}}q_m$ be all the 
transitions in $\delta$ from the state  $q$. Then the following
 properties hold.
\begin{enumerate}
\item For $0\leq i<n-1$, $distance_1({\tt d[i]},T(q),F_{\infty})\:=$ $\min\{x_1,...,x_m\}$
where \\
$x_j\:=$ $\max\{(1-simval(d_i,a_j)),distance_1({\tt d[i+1]},T(q_j),F_{\infty})\}$ for 
$j=1,...,m$.
\item If  there is at least one $j$ such that
$q_j\in Final$ then 
$distance_1(q,{\tt d[n-1]},F_{\infty})\:=$\\
$\min\{(1-simval(d_{n-1},a_j))\::q_j\in Final\}$, 
otherwise $distance_1(q,{\tt d[n-1]},F_{\infty})\:=\infty$.
\end{enumerate} 

{\bf Proof:} Part 1 of the lemma is seen as follows. Assume $i<n-1$ and observe that 
$T(q)\:=\bigcup_{1\leq j\leq m} (a_j T(q_j))$. Hence $distance_1({\tt d[i]},T(q),F_{\infty})\:=$
$\min \{distance_1({\tt d[i]}, a_j T(q_j), F_{\infty})\::1\leq j\leq m\}$. It should not
be difficult to see that for each $j$, $1\leq j\leq m$,  
$distance_1({\tt d[i]}, a_j T(q_j), F_{\infty})$ is $x_j$. Part 2 of the lemma
follows from the fact that 
the set of strings of unit length in $T(q)$ is the set $\{a_j\::q_j\in\:Final\}$.
$\Box$

It is easy to see that the complexity of this algorithm is $O(n\cdot Size({\cal A}))$. 
Thus the algorithm is of linear complexity in $n$ and in $Size({\cal A})$. 
A formal description of the algorithm is given in \cite{HS00}.

Now we show how to compute $distance_1(d,{\cal A}, F_{k})$
for any $k$ such that $0<k<\infty$. 
For a sequence $a$ over $\Delta \cup \{\phi\}$, let $elength(a)$ denote the number of 
values of $i$
such that $a_i\in \Delta$, i.e., $a_i$ is not the wild card symbol.
For any database sequence $d$ as given above and
any sequence $a=a_0,...,a_{n-1}$ over the alphabet $\Delta \cup \{\phi\}$ and for
any $k>0$, 
define an un-normalized distance $udist_k(d,a)$ as follows: 
$udist_k(d,a)\:=\Sigma_{0\leq i<n} (1-simval(d_i,a_i))^k$.
It should be easy to see that, if $elength(a)>0$ then $distance_1(d,a, F_{k})\:=$
$(\frac{udist_k(d,a)}{elength(a)})^{1/k}$; if $elength(a)=0$ then
$distance_1(d,a,F_{k})=0$.
For a database sequence $d$ and a language $L$ over $\Delta\cup \{\phi\}$, 
we define a set $Udist\_elength_k(d,L)$ of un-normalized distance and effective
length pairs as follows; $Udist\_elength_k(d,L)\:=$$\{(x,l)\::\exists a\in L$ such that
$elength(a)=l$ and $x$ is the minimum value of $udist_k(d,b)$ of all $b\in L$ whose
effective length is $l$, i.e., $elength(b)=l\}$.

The following lemma shows how $distance_1(d,{\cal A}, F_K)$ can be computed
from the sets $Udist\_elength_k({\tt d[0]},T(q))$ for each initial state $q$,
i.e., $q\in I$.

LEMMA 3.2: The following properties hold.
\begin{enumerate}
\item If the sets $Udist\_elength_k({\tt d[0]},T(q))$ for each initial state $q$
are all empty then $distance_1(d,{\cal A}, F_k)= 1$.
\item If $(0,0)\in Udist\_elength_k({\tt d[0]},T(q))$ for some $q\in I$ 
then $distance_1(d,{\cal A}, F_k)=0$.
\item If none of the above conditions holds then
 $distance_1(d,{\cal A}, F_k)=\:$$\min \{(\frac{x}{l})^{\frac{1}{k}}\::(x,l)$
$\in Udist\_elength_k({\tt d[0]},q)$ for some $q\in I\}$.
\end{enumerate}
{\bf Proof:} The condition of part 1 indicates that there are no
strings of the same length as $d$ that are accepted by ${\cal A}$ and hence
$distance_1(d,{\cal A}, F_k)= 1$. The condition of part 2 indicates that
there is a string containing only the symbol $\phi$ that is of the same length as
$d$ that is accepted by ${\cal A}$ and hence $distance_1(d,{\cal A}, F_k)=0$. 
Part 3 of the lemma follows from the definitions. \qed

The following lemma leads to a method for computing the sets
$Udist\_elength_k({\tt d[i]},T(q))$ for each state $q$ in decreasing values of $i$.

LEMMA 3.3: Let $q$ be any state in $Q$ and 
$q\rightarrow_{a_{1}}q_1,...,$$q\rightarrow_{a_{m}}q_m$ be all the 
transitions in $\delta$ from the state $q$. Then the following
 properties hold.
\begin{enumerate}
\item Let $i$ be an integer such that $0\leq i<n-1$ and 
$U\:=\{(x+(1-simval({\tt d[i]}, a_j))^k,\:l+1)\::$$1\leq j\leq m$ and $a_j\neq \phi$ and
$(x,l)\in Udist\_elength_k({\tt d[i+1]},T(q_j))\}\:\cup\:$
$\{(x,l)\in Udist\_elength_k({\tt d[i+1]}, T(q_j))\:: 1\leq j\leq m, a_j=\phi\}$.
Then, $Udist\_elength_k({\tt d[i]}, T(q))\:=$$\{(x,l)\in U\:: x$ is the minimum of all pairs of the form
$(y,l)\in U\}$.
\item If none of the $q_j$ is in $Final$ then 
$Udist\_elength_k({\tt d[n-1]},T(q))$ contains the single element $(\infty,1)$. 
Otherwise, 
if $\exists j$ such that $1\leq j\leq m$ and $a_j=\phi\:$ and $q_j\in Final$
then $Udist\_elength_k({\tt d[n-1]},T(q))$ contains the pair $(0,0)$;
if $\exists j$ such that $1\leq j\leq m$ and $a_j\neq \phi\:$ and $q_j\in Final$
then $Udist\_elength_k({\tt d[n-1]},T(q))$ contains the pair $(x,1)$
where $x\:= \min \{(1-\:simval({\tt d[n-1]},a_j))^k \::a_j\neq \phi$ and $q_j\in Final\}$.
\end{enumerate}

{\bf Proof:} Part 1 of the lemma follows from the definition of $Udist\_elength_k({\tt d[i]}, T(q))$
and the fact that $T(q)\:=\cup_{1\leq j\leq m} a_j T(q_j)$. Part 2 of the lemma follows from the
observations. If none of the $q_j$ is in $Final$ then $T(q)$ has no strings of length 1 and hence
$Udist\_elength_k({\tt d[n-1]},T(q))$ contains the single element $(\infty,1)$.
If $\exists j$ such that $1\leq j\leq m$ and $a_j=\phi\:$ and $q_j\in Final$
then the string $\phi$ of length $1$ is in $T(q)$ and hence $(0,0)\in Udist\_elength_k({\tt d[n-1]},T(q))$.
If $\exists j$ such that $1\leq j\leq m$ and $a_j\neq \phi\:$ and $q_j\in Final$
then from the definitions it is seen that $(x,1)\in Udist\_elength_k({\tt d[n-1]},T(q))$
where $x$ is as given in the lemma. It is to be noted that $Udist\_elength_k({\tt d[n-1]},T(q))$
contains at most two elements. $\Box$

It is to be noted that for any element $(x,l)$ in $Udist\_elength_k({\tt d[i]}, T(q))$,
$0\leq l\leq n-i$. Further more, for any two elements $(x,l)$ and $(x',l')$ in 
$Udist\_elength_k({\tt d[i]}, T(q))$ it is the case that $l\neq l'$. As a consequence,
$Udist\_elength_k({\tt d[i]}, T(q))$, has at most $n-i+1$ elements. Using part 2 of the above
lemma, we see that  the values of of the set $\{Udist\_elength_k({\tt d[n-1]},T(q))\::q\in Q\}$
can all be computed in time $O(Size({\cal A}))$. Using part 2, of the lemma, we see that
the values in the set $\{Udist\_elength_k({\tt d[i]},T(q))\::q\in Q\}$ can be computed
from the values in the set $\{Udist\_elength_k({\tt d[i+1]},T(q))\::q\in Q\}$ in time
$O((n-i)\cdot Size({\cal A}))$. Using the last step repeatedly, we see that the values
in the set $\{Udist\_elength_k({\tt d[0]},T(q))\::q\in Q\}$ can be computed
in time $O(n^2\cdot Size({\cal A}))$. Hence $distance_1(d,{\cal A}, F_k)$ can be computed
in time $O(n^2\cdot Size({\cal A}))$ where $n=length(d)$. Thus the algorithm is of complexity
quadratic in the length of $d$ and linear in $Size({\cal A})$.

If the input symbol $\phi$ does not appear in any string in the language $L({\cal A})$ then
we can delete all transitions on the symbol $\phi$ from the transition set $\delta$
of the automaton ${\cal A}$. The resulting automaton ${\cal A}$ has no transitions on $\phi$.
In this case, it is not difficult to
see that, for each $i=0,..,n-1$, the set $Udist\_elength_k({\tt d[i]},T(q))$ has at most one element.
Hence, the set of values $\{Udist\_elength_k({\tt d[i]},T(q))\::q\in Q\}$ can be computed
from the values in the set $\{Udist\_elength_k({\tt d[i+1]},T(q))\::q\in Q\}$ in time $O(Size({\cal A}))$
only. As a consequence the set of values $\{Udist\_elength_k({\tt d[0]},T(q))\::q\in Q\}$ can be computed
in time $O(n\cdot Size({cal A}))$. Hence $distance_1(d,{\cal A}, F_k)$ can be computed in time
$O(n\cdot Size({\cal A}))$. Thus, the resulting algorithm is only of linear complexity
in the length of $d$ as well as in $Size({\cal A})$.

\noindent{\bf Computing the value of $distance_2(d,{\cal A}, F)$}

The proof of the following lemma gives a method for computing $distance_2(d,{\cal A},F)$
where $F$ is any of the vector distance functions given previously. The complexity of the algorithm
is triple exponential in the number of states of ${\cal A}$, and linear or quadratic in the
length of $d$.

LEMMA 3.4: For a database sequence $d$ of length $n$, automaton ${\cal A}$ with $m$ number of states
 and vector distance function $F$,
there exists an algorithm that computes $distance_2(d,{\cal A}, F)$ which is of complexity
$O(2^{2^{2^{2m}}}\cdot p(n))$ where 
$p(n)$ is $n$ if $F\:= F_{\infty}$, and is $n^2$ if $F=F_i$ for $i<\infty$.

{\bf Proof:} We prove the lemma by giving an algorithm, of the appropriate complexity, that computes 
$distance_2(d,{\cal A},F)$.
From the definition, we have 
$distance_2(d,{\cal A},F)\:=$ $distance_1(d, maximal(closure(L({\cal A}))),F)$.
Let $\Delta'\:= \Delta \cup \{\phi\}$ where $\phi$ is the wild card symbol.
Recall that the elements of $closure(L({\cal A}))$ are strings over the alphabet $\Delta'$.
We construct an automaton  ${\cal H}$ that accepts the language $maximal(closure(L{\cal A}))$.
Then we simply compute $distance_2 (d,{\cal A}, F)$ to be the value of $distance_1(d, {\cal H}, F)$.

Now we show how to compute the automaton ${\cal H}$. First we compute the automaton $\overline{{\cal A}}$
which accepts the complement of the language accepted by the automaton ${\cal A}$. 
From this automaton, we construct another automaton ${\cal C}$ which accepts the complement
of the language $closure(L({\cal A}))$, i.e., the language $\overline{closure(L({\cal A}))}$. 
The construction of ${\cal C}$ uses the following fact.
If a string $\alpha \in (\Delta')^*$ is in $\overline{closure(L({\cal A}))}$ then
there exists another string $\beta\in \Delta^*$ obtained from $\alpha$ by replacing each occurrence
of $\phi$ in it with some symbol in $\Delta$ such that $\beta\in \overline{L ({\cal A})}$.
${\cal C}$ simulates $\overline{{\cal A}}$ on an input string
with the following modification. 
Whenever it sees the input symbol $\phi$, it replaces it, non-deterministically,
by some symbol from $\Delta$ and simulates $\overline{{\cal A}}$ on the guessed symbol.
It accepts it if $\overline{{\cal A}}$ accepts. It is not difficult to see
that ${\cal C}$ accepts a string $\alpha$ over the alphabet $\Delta'$ iff there exists
a string $\beta$, obtained by replacing every occurrence of the $\phi$ symbols in $\alpha$  by some symbol from $\Delta$,
which is accepted by $\overline{{\cal A}}$. Hence, it is easy to see that ${\cal C}$ accepts the
the language $\overline{closure(L({\cal A}))}$. It is not difficult to see that 
we can obtain such an automaton ${\cal C}$ such that $|{\cal C}|\:=|\overline{{\cal A}}|$.

Next we construct
the automaton $\overline{{\cal C}}$ which accepts the complement of the 
language accepted by ${\cal C}$, i.e., which accepts the language $closure(L({\cal A}))$.
It is to be noted that $|\overline{{\cal C}}|\leq 2^{|{\cal C}|}$ 
and hence $|\overline{{\cal C}}|\leq 2^{|\overline{{\cal A}}|}$. 
Similarly,  $|\overline{{\cal A}}|\leq 2^{|{\cal A}|}$.
From this we see that $|\overline{{\cal C}}| \leq 2^{2^{|{\cal A}|}}$
and hence $|\overline{{\cal C}}| \leq 2^{2^{m}}$.

Using $\overline{{\cal C}}$, we construct an automaton ${\cal D}$ which accepts the complement
of the language $maximal(closure(L({\cal A}))$. 
The automaton ${\cal D}$ on an input string $\alpha$ over $\Delta'$ acts as follows.
It accepts $\alpha$ if either $\alpha \notin closure(L({\cal A}))$, or  there exists another string 
$\beta \in closure(L({\cal A}))$ such that $\alpha < \beta$.
The former condition is checked
by simulating ${\cal C}$ over $\alpha$. To check the later condition, ${\cal D}$ non-deterministically
changes at least one of the input symbols in $\sigma$, which is an element of $\Delta$, to 
 $\phi$ and checks that the resulting string is accepted by $\overline{{\cal C}}$, i.e., is
in $closure(L({\cal A}))$. It is easy to see that we can construct such an automaton ${\cal D}$
such that its number of states is linear in the number of states of $\overline{{\cal C}}$, 
and hence is
double exponential in $m$. Next we construct the complement $\overline{{\cal D}}$
 of ${\cal D}$. Clearly $\overline{{\cal D}}$ accepts the language $maximal(closure({\cal A}))$.
We take ${\cal H}$ to be the automaton $\overline{{\cal D}}$.
Clearly, $|{\cal H}|\leq 2^{2^{2^{m}}}$. Note that $Size({\cal H})$ which is the 
sum of its number of states and transitions is quadratic in $|{\cal H}|$.
Hence $Size({\cal H})\leq 2^{2^{2^{2m}}}$. 

For each $i=1,...,\infty$, we compute $distance_2(d,{\cal A}, F_i)$ to be 
$distance_1(d,{\cal H}, F_i)$. For $i=\infty$, the complexity of the algorithm is
$O(n\cdot Size({\cal H}))$ and hence is $O(n\cdot 2^{2^{2^{2m}}})$.  .
For $i<\infty$, the complexity is $O(n^2\cdot Size({\cal H}))$
and hence is $O(n^2 \cdot 2^{2^{2^{2m}}})$.   
$\Box$

\section{Temporal Logic}
\label{temp-sec}

In this section, we consider linear Temporal logics as one of the formalism
for specifying queries over database sequences.
Such logics have been extensively used in specification of 
properties of concurrent programs \cite{MP92}. They have also been used 
in database systems for specifying queries in Temporal databases 
\cite{Cho92a,Cho92b}
and for specifying triggers in active database systems \cite{SW95a,SW95b}.
We assume that we have a finite set ${\cal P}$ whose members are called
atomic propositions. 
Each member of this set denotes an atomic predicate over a database state.
Formulas of Temporal Logics (TL) are formed from atomic propositions 
using the propositional connectives $\wedge,\vee, \neg$ and the temporal
operators $\X$ (``nexttime'') and $\U$ (``until''). 
The set of formulas of TL is the smallest set satisfying the following
conditions. Every atomic proposition is a formula of TL; both $true$ and
$false$ are formulas; if 
$g$ and $h$ are formulas of TL then $g\wedge h$, $g\vee h$, $\neg g$, $\X g$
and $g\U h$ are also formulas of TL. 
For a formula $f$, we let $length(f)$ denote its length.

Given a database sequence $d=(d_0,d_1,...,d_{n-1})$ and a temporal 
formula $f$, and a vector distance function $F$,
we define a distance function $syndist(d,f,F)$ inductively based
on the syntax of $f$ as follows.
\begin{itemize}
\item For an atomic proposition $P$, $syndist(d,P,F)\:=\:1-simval(d_0,P)$.
\item $syndist(d, g\wedge h,F)\:=\:\max \{syndist(d,g,F), syndist(d,h,F)\}$.
\item $syndist(d, g\vee h,F)\:=\:\min \{syndist(d,g,F), syndist(d,h,F)\}$.
\item $syndist(d, \neg g)\:=\: 1 - syndist(d,g,F)$.
\item $syndist(d,\X g,F)\:=\:syndist({\tt d[1]},g,F)$ if $length(d)>1$; otherwise, $syndist(d,\X g,F)=\infty$.
\item $syndist(d, g\U h,F)\:=\:\min \{F(\vec{U_i},\vec{1})\::0\leq i<n\}$ where 
$\vec{U_i}$ is the vector $(u_{i,0},u_{i,1},..,u_{i,i})$ whose components
are given as follows.
 $u_{i,i}\:=1-syndist({\tt d[i]},h,F)$ and for $j$, $0\leq j<i$, 
$u_{i,j}\:= 1-syndist({\tt d[j]},g,F)$. Intuitively, this definition corresponds to
the exact semantics of $\U$.  
\end{itemize}

Now, we define two types of semantic distance functions between a database 
sequence $d$ and a TL formula $f$. To do this, we need the following
definitions.
Let $\Delta$ be the set of all subsets of atomic propositions, i.e.
$\Delta\:=\:2^{{\cal P}}$.
Let $s\:=\:(s_0,...,s_{n-1})$ be any sequence over $\Delta$.
Now we define the satisfaction of $f$ at the beginning of $s$ inductively
on the structure of $f$ as follows.\\
\begin{itemize}
\item For an atomic proposition $P\in {\cal P}$, $s$ satisfies 
$P$ if $P\in s_0$.
\item $s$ satisfies $g\wedge h$ if $s$ satisfies both $g$ and $h$.
$s$ satisfies $g\vee h$ if $s$ satisfies either $g$ or $h$.
\item $s$ satisfies $\neg g$ if $s$ does not satisfy $g$.
\item $s$ satisfies $\X g$ if $n>0$ and ${\tt s[1]}$ satisfies $g$.
\item $s$ satisfies $g\U h$ if there exists an $i<n$ such that
${\tt s[i]}$ satisfies $h$ and for all $j$, $0\leq j<i$, ${\tt s[j]}$ satisfies
$g$.
\end{itemize}
 
We say that two TL formulas $f$ and $g$ are equivalent if the sets of
sequences (over $\Delta$) that satisfy them are identical.
Let $f$ be a TL formula and  ${\cal P}\:=\{P_0,P_1,...,P_{m-1}\}$
be the set of atomic propositions that appear in $f$. Let
$s\:=(s_0,...,s_{n-1})$ be any sequence over $\Delta$. Let
$\Psi$ be the set consisting of the elements of ${\cal P}$ and
negations of elements in ${\cal P}$. Formally,
 $\Psi\:={\cal P} \cup \{\neg P_i\:: 0\leq i<m \}$. 
Now we define a sequence $expn(s)$  over $\Psi$ which is obtained from
$s$ by expanding each $s_i$ in to a subsequence of length $m$ whose $j^{th}$
element is $P_j$ or $\neg P_j$ depending on whether $P_j$ is in $s_i$ or not.
Formally, $expn(s)\:=(t_0,...,t_{nm-1})$ is a sequence of length 
$nm$ defined as follows:
 for each $i,j$ such that $0\leq i<n$ and
$0\leq j<m$,  if $P_j\in s_i$ then $t_{im+j}=P_j$, otherwise $t_{im+j}=\neg P_j$.

For a TL formula $f$, let $L(f)\:=\{expn(s):\:s$ satisfies $f\}$.
Note that $L(f)$ is a language over $\Psi$. For any positive integer
$r$, let $L_r(f)\:=\{expn(s):\:length(s)=r$ and $s$ satisfies $f\}$.
$L_r(f)$ corresponds to those sequences of length $r$ that satisfy
$f$.

Let $d=(d_0,d_1,...,d_{n-1})$ be a database sequence.
Now, we define another database sequence $expn(d)$ obtained from $d$
by repeating each $d_i$ successively $m$ times
(recall $m$ is the number of atomic propositions), i.e.,
$expn(d)\:=((d_0)^m,(d_1)^m,...,(d_i)^m,...,(d_{n-1})^m))$.
Let $f$ be a TL formula and $F$ be a vector distance function.
Now, for each $j=1,2$, we define a semantic distance
of $d$ with respect to $f$ and $F$ (denoted by $semdist_j(d,f,F)$)
as follows: $semdist_j(d,f,F)\:=distance_j(expn(d), L_n(f), F)$.
Recall that $distance_j$ is defined in the previous subsection.

It is to be noted that we have assumed the set ${\cal P}$ to be exactly the set
of atomic propositions that appear in $f$. However, if we take ${\cal P}$ to
be any super set of the set of atomic propositions appearing in $f$, then it is
easy to see that the syntactic distances, i.e., $syndist(d,f,F_k)$ for $k$ such that
$1\leq k\leq \infty$, remain the same. It can be shown that similarly
$semdist_2(d,f,F_k)$ for all $k=1,...,\infty$  and 
$semdist_1(d,f,F_\infty)$ remain the same. That is all these distance measures
depend only on the similarity values of atomic propositions that appear
in $f$ and not on other atomic propositions.
On the other hand, this property does not hold for the distance measures
$distance_1(d,f,F_k)$ for $k\neq \infty$. 

The semantic distances of a database sequence with respect to equivalent
TL formulas are equal (i.e., if $f$ and $g$ are equivalent then 
$semdist_j(d,f,F)\:=semdist_j(d,g,F)$).
However this property does not hold for syntactic distances.
For example, the syntactic distance of a database sequence with
respect to the two equivalent formulas 
$(P\wedge Q)\vee (P\wedge \neg Q)$  and $P$ may be different.
The following lemma shows that $syndist(d,f,F_{\infty})\leq semdist_2(d,f,F_{\infty})$.
The lemma can be proven by induction on the structure of the formula $f$.

LEMMA 4.1: For any database sequence $d$  and TL formula $f$ in which all negations are applied
to atomic propositions,
$syndist(d,f,F_{\infty})\leq semdist_1(d,f,F_{\infty})$.  

{\bf Proof:} The proof has two steps. In the firs step, we 
show that any formula $g$, in which all negations are applied only to atomic propositions,
can be transformed to a formula $G(g)$, that has no $\U$ operator appearing in it and in which
all negations appear only to atomic propositions, such that 
$syndist(d,g,F_{\infty})\:=syndist(d,G(g),F_{\infty})$ and
$semdist_1(d,g,F_{\infty})\:= semdist_1(d,G(g),F_{\infty})$.
Let $n = length(d)$.
The formula $G(g)$ is defined inductively on the structure of $g$ as follows.
If $g$ is an atomic proposition or the negation of an atomic proposition
then $G(g)=g$. If $g\:=g_1\wedge g_2$ or $g\:=g_1\vee g_2$ or $g\:=\X g_1$ then 
$G(g)\:=G(g_1) \wedge G(g_2)$ or $G(g)\:=G(g_1) \vee G(g_2)$ or $G(g)\:=\X G(g_1)$
, respectively. If $g\:=g_1\U g_2$ then 
$G(g)\:= \bigvee_{0\leq i<n}$ $( (\bigwedge_{0\leq j<i} (\X)^j g_1) \wedge (\X)^i (g_2))$.
In the above definition $(\X)^j$ denotes a string of $j$, $\X$ operators. 
It is to be noted that in the definition of $G(g)$ for $g\:=g_1\U g_2$, we are replacing the 
$\U$ operator by $n$ disjuncts (recall that $n=length(d)$); the $i^{th}$ disjunct asserts 
that $g_2$ is satisfied after
$i$ database states and at all the intermediate states $g_1$ is satisfied; 
also note that the $i^{th}$ disjunct has $i$ conjuncts.
By a simple induction on the structure of $g$, one can easily show that
$syndist(d,g,F_\infty)\:=syndist(d,G(g),F_\infty)$. It is to be noted that
this property is not satisfied if we replace $F_\infty$ by any other $F_k$.
It can also be shown that for any sequence $s$ of length $n$ over $\Delta$,
$s$ satisfies $g$ iff $s$ satisfies $G(g)$; i.e., the set of sequences of length
$n$ that satisfy $g$ is same as the set of sequences of length $n$ that satisfy $G(g)$.
As a consequence, $semdist_1(d,g,F_\infty)\:=semdist_1(d,G(g),F_\infty)$.

Now we show, by induction, that for any formula $g$ that does not contain any $\U$ operator
and in which all negations are applied only to atomic propositions,
$syndist(d,g,F_{\infty})\leq semdist_1(d,g,F_{\infty})$.  
For the base case, i.e., when $g$ is an atomic proposition or the negation of an atomic proposition
the property trivially holds. Now consider the cases when $g\:=g_1 \vee g_2$, or $g\:=g_1 \wedge g_2$.
As induction hypothesis, assume $syndist(d,g_1,F_{\infty})\leq semdist_1(d,g_1,F_{\infty})$
and  $syndist(d,g_2,F_{\infty})\leq semdist_1(d,g_2,F_{\infty})$.
Observe that $\min \{syndist(d,g_1,F_\infty),\:syndist(d,g_2,F_\infty)\}$ 
$\leq \min \{semdist_1(d,g_1,F_\infty),\:semdist_1(d,g_2,F_\infty)\}$.
From the definitions, we see that the left hand side of the above inequality is
$syndist(d,g_1\vee g_2,F_\infty)$ and the right hand side equals   
$semdist_1(d,g_1\vee g_2,F_\infty)\}$. Hence, it is the case that
$syndist(d,(g_1\vee g_2), F_{\infty})$$\leq semdist_1(d,(g_1\vee g_2), F_{\infty})$.
It is also easy to see that
$\max \{syndist(d,g_1,F_\infty),\:syndist(d,g_2,F_\infty)\}$ 
$\leq \max \{semdist_1(d,g_1,F_\infty),\:semdist_1(d,g_2,F_\infty)\}$.
The left hand side of this inequality is 
$syndist(d,(g_1\wedge g_2), F_{\infty})$.
We show that its right hand side is less than or equal to
$semdist_1(d,(g_1\wedge g_2),F_\infty)$. Let $X_1\:=L_n(g_1)$
and $X_2\:=L_n(g_2)$. From the definitions, we have
$semdist_1(d,(g_1\wedge g_2),F_\infty)$
$\:=\min \{dist(expn(d),s,F_\infty)\:: s\in X_1\cap X_2 \}$,
$semdist_1(d,g_1,F_\infty)$
$\:=\min \{dist(expn(d),s,F_\infty)\:: s\in X_1 \}$, and
$semdist_1(d,g_2,F_\infty)$
$\:=\min \{dist(expn(d),s,F_\infty)\:: s\in X_2 \}$.
From these we see that both $semdist_1(d,g_1,F_\infty)$ and
$semdist_1(d,g_2,F_\infty)$ are less than or equal to
$semdist_1(d,(g_1\wedge g_2),F_\infty)$.
Hence $\max \{semdist_1(d,g_1,F_\infty),\:semdist_1(d,g_2,F_\infty)\}$
is less than or equal to $semdist_1(d,g_1\wedge g_2,F_\infty)$.

Now consider the case when $g=\:\X g_1$.
If $length(d)\leq 1$ then $syndist(d,g,F_\infty)=\infty$, and in this case
$semdist_1(d,g,F_\infty)$ is also $\infty$ because all strings that satisfy $g$
are of length at least two. Now consider the case when $length(d)\geq 2$ 
and as induction hypothesis assume that
$syndist({\tt d[1]},g_1,F_\infty)\leq semdist_1({\tt d[1]},g_1,F_\infty)$.
From the definitions, we see that $syndist(d,g,F_\infty)$ equals
the left hand side of this inequality. We show that the right hand
side is less than or equal to $semdist_1(d,g,F_\infty)$ and from this
it would follow that $syndist(d,g,F_\infty)\leq semdist_1(d,g,F_\infty)$.
Let $G_1\:=L_{n-1}(g_1)$ and $G\:=L_n(g)$.
From the definition of $L_n(g)$, we see that
$G\:=\{ expn(\delta) t\::\delta\in \Delta$ and $t \in G_1\}$.
By definition 
$semdist_1({\tt d[1]},g_1,F_\infty)\:=\min \{dist(expn(d),u,F_\infty)\::u\in G_1\}$
and 
$semdist_1(d,g,F_\infty)\:=\min \{dist(expn(d),u,F_\infty)\::u\in G\}$.
Let $s\in G_1$ be the string such that 
$dist(expn({\tt d[1]}),s,F_\infty)\:= \min \{dist(expn({\tt d[1]}),u,F_\infty)\::u\in G_1\}$.
Now, it is not difficult to see that
$semdist_1(d,g,F_\infty)\:=\min \{dist(expn(d),expn(\delta)s,F_\infty)\::\delta\in \Delta\}$
and hence $semdist_1(d,g,F_\infty)\geq semdist_1({\tt d[1]},g_1,F_\infty)$.
$\Box$
 
It is to be noted that $syndist(d,f,F_i)\leq syndist(d,f,F_j)$ for all $i,j$
such that $1\leq i\leq j\leq \infty$. Also $semdist_2(d,f,F_\infty)\leq semdist_1(d,f,F_\infty)$
and $semdist_1(d,f,F_i)\leq semdist_1(d,f,F_j)$ and $semdist_2(d,f,F_i)\leq semdist_2(d,f,F_j)$
for all $i,j$ such that $i\leq j\leq \infty$. These results follow directly from lemma 2.1.
The only known non-trivial relationship between syntactic and semantic distances for temporal formulas
is the one given by lemma 4.1. For example, in general, we believe that neither the relation
$syndist(d,f,F_\infty)\leq semdist_2(d,f,F_\infty)$ nor the reverse relationship 
holds. Similarly, in general, for $i<\infty$, we can not relate
$syndist(d,f,F_i)$ with  either $semdist_1(d,f,F_i)$ or $semdist_2(d,f,F_i)$.

{\bf Algorithm for computing the Syntactic distance}

Now we present algorithms for computing the syntactic and the semantic 
distances. First we present the algorithms for computing the
syntactic distances. 

LEMMA 4.2: Given a database sequence $d\:=(d_0,...,d_{n-1})$ and a TL formula
$f$ and given the similarity values of the database
states in $d$ with respect to the atomic propositions
appearing in $f$, there exists an algorithm that computes
 $syndist(d,f,F_\infty)$ in time $O(n\cdot length(f))$.
For each $k$, $0<k<\infty$, there exists an algorithm that computes
$syndist(d,f,F_k)$ in time $O(n^2 \cdot length(f))$.

{\bf Proof:} Let $d,f$ be as given in the lemma. For each
$i$, $0\leq i<n$ and for each $g$
which is an atomic proposition or its negation, let
$simval(d_i,g)$ be the similarity value of $g$ in the database state $d_i$. 
Let $SF(f)$ be  the set of all sub-formulas of $f$. Let $k$ be any integer
such that $0<k\leq \infty$. 
For each $g\in SF(f)$ and for each $i=0,...,n-1$, we compute
$syndist({\tt d[i]},g,F_k)$ inductively on the length of $g$ as follows.
The algorithms for computing $syndist({\tt d[i]},g,F_k)$ is same for all $k$
in all cases excepting in the case when $g$ is of the form $g_1\U g_2$.
\begin{itemize}

\item When $g$ is an atomic proposition or its negation, $syndist({\tt d[i]},g,F_k)\:=1-simval(d_i,g)$.

\item When $g=g_1\wedge g_2$, we compute $syndist({\tt d[i]},g,F_k)$ to be the maximum
of $syndist({\tt d[i]},g_1,F_k)$ and $syndist({\tt d[i]},g_2,F_k)$. 
\item When $g=g_1\vee g_2$,
then we compute $syndist({\tt d[i]},g,F_k)$ to be the minimum of 
$syndist({\tt d[i]},g_1,F_k)$ and $syndist({\tt d[i]},g_2,F_k)$. 
\item When $g= \X g_1$,
$syndist({\tt d[i]},g,F_k)$ is taken to be $1$ for $i=n-1$, and
it is taken to be $syndist({\tt d[i+1]},g_1,F_k)$ for $i<n-1$.
\item For the case when $g\:=g_1\U g_2$ we do as follows..
We compute $syndist({\tt d[i]},g,F_k)$ for decreasing values
of $i$. We first give the method for the case when $k=\infty$.
The value of $syndist({\tt d[n-1]},g,F_\infty)$
 is computed to be $syndist({\tt d[n-1]},g_2,F_{\infty})$.
For $i<n-1$, $syndist({\tt d[i]},g,F_{\infty})$ is computed to be the minimum
of the two values--- $syndist({\tt d[i]},g_2,F_{\infty})$ and 
$\max \{syndist({\tt d[i]},g_1,F_{\infty}),\:syndist({\tt d[i+1]},g,F_{\infty})\}$.
It is easy to see that this procedure only takes $O(n)$ time.

For $k\neq \infty$, we compute the values $\{syndist({\tt d[i]},g,F_k):0\leq i<n\}$
from the values $\{syndist({\tt d[i]},g_1,F_k), syndist({\tt d[i]},g_2,F_k)\:: 0\leq i<n\}$
as follows. Let $sum_{i,j}$ be the sum of $(syndist({\tt d[r]},g_1,F_k))^k$ for all values of
$r$ such that $i\leq r<j$. Let $y_{i,j}\:= (\frac{sum_{i,j}+syndist({\tt d[j]}, g_2, F_k)}{j-i+1})^{\frac{1}{k}}$.
We compute $syndist({\tt d[i]},g,F_k)$ to be minimum of the values $y_{i,j}$ for
$j=i,...,n-1$. From the definitions it is not difficult to see that this procedure
correctly computes $syndist({\tt d[i]},g, F_k)$. It is not difficult
to see that all the values of $sum_{i,j}$ and $y_{i,j}$ for $j=i,...,n-1$ can be computed
in time $O(n)$. Thus this step for computing $syndist({\tt d[i]},g,F_k)$ takes $O(n)$ time.
Computing all the values in $\{syndist({\tt d[i]},g,F_k):0\leq i<n\}$ takes $O(n^2)$ time.

\end{itemize}

It is easy to see that the above algorithm correctly computes the syntactic
distances. The complexity of the  algorithm is $O(length(d)\cdot length(f))$ for
$k=\infty$, and for all cases when $k<\infty$
the complexity of the algorithm is $O(length(d)^2\cdot length(f))$. $\Box$

{\bf Computing the Semantic Distances}

Let $d\:=(d_0,d_1,...,d_{n-1})$ be a database sequence and $f$ be a
TL formula. 
We take an automata theoretic approach for computing $semdist_k(d,f,F_i)$ for
$i=1,...,\infty$. For this we use the well known result (see \cite{VWS83, ES83}) 
that shows that there exists an automaton ${\cal A}$ such that $|{\cal A}|\leq 2^{length(f)}$
and $L({\cal A})\:=L(f)$. The value of $symdist_1(d,f,F_i)$ is
computed as the value $distance_1(d,{\cal A}, F_i)$ using the algorithm given in
section 3.
 
To compute the value of $semdist_2(d,f,F_i)$ also, we use the approach 
given in section \ref{aut-sec}.
 This approach uses the complement $\overline{{\cal A}}$ of the automaton
${\cal A}$. The automaton $\overline{{\cal A}}$ accepts the language $\overline{L({\cal A})}$,
i.e., the complement of the language $L({\cal A})$; this is exactly the set of sequences
that satisfy the formula $\neg f$. Thus we can take $\overline{{\cal A}}$ to be the
automaton that accepts the set of all sequences that satisfy $\neg f$. Using the approach given
in \cite{VWS83,ES83} we can obtain such an automaton whose number of states is $O(2^{length(f)})$.
Using this automaton, we can apply the procedure given in section \ref{aut-sec}.
The complexity of the resulting algorithm will be $O(length(d)\cdot 2^{2^{2\cdot length(f)}})$ for the case when we
use the distance function $F_{\infty}$; for all other distance functions $F_i$ ($i<\infty$),
the complexity is $O(length(d)^2 \cdot 2^{2^{2\cdot length(f)}})$.

\section{Regular Expressions}
\label{reg-sec}

In this section we consider regular expressions (REs) as query languages
and define syntactic and semantic distances of a database sequence with
respect to REs. Let $\Delta$ be a finite set of atomic
queries. The set of REs over $\Delta$ is the smallest set of
strings satisfying the following conditions. Every element of $\Delta$
is a RE; if $g$ and $h$ are REs then $(g\vee h), gh$ and $(g)^*$ are
also REs. With each RE $f$ over $\Delta$, we associate a language $L(f)$
over $\Delta$ defined inductively as follows. For every $a\in \Delta$,
$L(a)\:=\{a\}$. For the REs $g,h$, $L(gh)\:=L(g)L(h)$, 
$L(g\vee h)\:= L(g)\cup L(h)$, $L((g)^*)\:=(L(g))^*$.

Let $d=(d_0,...,d_{n-1})$ be a database sequence and $f$ be a RE. 
For each $k=1,...,\infty$
we define a syntactic distance function, denoted by $syndist(d,f,F_k)$
inductively on the structure of $f$ as follows.
First we define this for the case when $k\neq \infty$.
\begin{itemize}
\item For $a\in \Delta$, $syndist(d,a,F_k)\:=1-simval(d_0,a)$ when $n=1$, i.e.
$length(d)=1$; otherwise $syndist(d,a,F_k)\:=\infty$.
\item $syndist(d,g\vee h,F_k)\:=\min \{syndist(d,g,F_k),\:syndist(d,h,F_k)\}$.
\item $syndist(d,gh,F_k)\:= \min \{ (\frac{length(\alpha)\cdot u^k+length(\beta)\cdot v^k}{n})^{\frac{1}{k}}\::$ 
$\alpha,\:\beta$ are, possibly null, database sequences such that $\alpha\beta\:=d$ and
$u=syndist(\alpha,g,F_k)$ and $v=syndist(\beta,h,F_k)\}$.
\item We define $syndist(d,(g)^*,F_k)$ as follows. 
If $d$ is the null string then  $syndist(d,(g)^*,F_k)=0$, otherwise, 
$syndist(d,(g)^*,F_k)$
is $\min \{(\frac{u_1+...+u_l}{n})^{\frac{1}{k}}\::$ 
$\exists \alpha_1,...,\alpha_l$ such that 
$\alpha_1\alpha_2...\alpha_l=d$ and for $1\leq i\leq l$,  $\alpha_i$ is 
non-null and $u_i\:=syndist(\alpha_i,g,F_k) \}$
\end{itemize}

The values of $syndist(d,f,F_{\infty})$ are defined as follows.
\begin{itemize}
\item For $a\in \Delta$, $syndist(d,a,F_{\infty})\:=1-simval(d_0,a)$ when $n=1$, i.e.
$length(d)=1$; otherwise $syndist(d,a,F_{\infty})\:=\infty$.
\item For the RE $g\vee h$, $syndist(d,g\vee h,F_{\infty})\:=
\min \{syndist(d,g,F_{\infty}),$$\:syndist(d,h,F_{\infty})\}$.
\item $syndist(d,gh,F_{\infty})\:= \min \{ \max\{syndist(\alpha,g,F_{\infty}),\:
syndist(\beta,g,F_{\infty})\}\::$
$\alpha,\:\beta$ are database sequences such that 
$\alpha\beta\:=d$ $\}$.
\item We define $syndist(d,(g)^*,F_{\infty})$ as follows. We define $syndist(d,(g)^*,F_{\infty})$
to be $\min \{ \max \{syndist(\alpha_1,g,F_{\infty}),...,syndist(\alpha_l,g,F_{\infty})\}\::$
$\alpha_1\alpha_2...\alpha_l=d$ and each $\alpha_i$ is non-null$\}$
\end{itemize}

For each $k\:=1,...,\infty$, we define two semantic distance functions
$semdist_{1}$ and $semdist_{2}$ as follows.
For any database sequences $d$ and RE $f$ and for $j=1,2$, 
$semdist_{j}(d,f,F_k)\:=distance_j(d,L(f),F_k)$.
The following lemma can be easily proven. It shows that the syntactic distance
and the semantic distance given by $semdist_{1}$ are identical.

LEMMA 5.1: For each databases sequence $d$ and RE $f$ and for each 
$k=1,...,\infty$, $syndist(d,f,F_k)\:=semdist_{1}(d,f,F_k)$.  

{\bf Proof:} We give the proof for the case when $k<\infty$. The proof is similar
for the case $k\:=\infty$.
The lemma is proved by induction on the length of $f$.
In the base case, the length of $f$ is one and 
$f\:=a$ for some $a\in \Delta$. From definition of the two distance measures, 
it is easy to see that $syndist(d,f,F_k)\:=semdist_{1}(d,f,F_k)$.  
As an induction hypothesis, assume that the lemma is true for all
$f$ of length less than or equal to $r$. Now consider a RE $f$ whose length
is $r+1$. We consider the different cases. The first case is when
$f$ is of the form $g\vee h$. In this case, from the definitions,
we have $semdist_1(d,g\vee h,F_k)\:= distance_1(d,L(g\vee h),F_k)$
and which equals $distance_1(d,L(g)\cup L(h),F_k)$, which is
$\min\{distance_1(d,L(g),F_k),\:distance_1(d,L(h),F_k)\}$. The later
is  $\min\{semdist_1(d,g,F_k),\:semdist_1(d,h,F_k)\}$ and this 
by induction equals $\min \{syndist(d,g,F_k),syndist(d,h,F_k)\}$
which is $syndist(d,f,F_k)$.
 
Now consider the case when $f=gh$. We have 
$semdist_1(d,gh,F_k)\:= distance_1(d, L(g)L(h),F_k)$.
The later is $\min \{distance_1(d,\alpha\beta,F_k)\::\alpha \in L(g),\:\beta \in L(h)\}$.
This value is given by \\
(*) $\min \{distance_1(d_1d_2,\alpha\beta,F_k)\::\alpha \in L(g),\:\beta \in L(h),$$
length(d_1)=length(\alpha), d_1d_2=d\}$.\\
It is easy to see that, for database sequences $d_1,d_2$ such that $d=d_1d_2$ and $length(\alpha)=length(d_1)$
and $length(d)=length(\alpha\beta)$, 
$distance_1(d_1d_2,\alpha\beta,F_k)\:=$$(\frac{length(d_1)\cdot u^k + length(d_2)\cdot v^k}{length(d)})^{\frac{1}{k}}$
where $u=distance_1(d_1,\alpha,F_k)$ and $v=distance_1(d_2,\beta,F_k)$.
Let $E$ denote the expression $(\frac{length(d_1)\cdot u^k + length(d_2)\cdot v^k}{length(d)})^{\frac{1}{k}}$. Substituting this in (*), we get 
$semdist_1(d,gh,F_k)\:=\min\{\:E\::\alpha \in L(g),\:\beta \in L(h),$$
length(d_1)=length(\alpha), d_1d_2=d\}$. Since we are taking the minimum on the right hand side,
it is not difficult to see that we can choose $\alpha$ to be the one that gives the minimum
value for $u$ and this minimum value of $u$ is $semdist_1(d_1,g,F_k)$. Similarly, we take
$v$ to be the  $semdist_1(d_2,h,F_k)$. Thus we get,
 $semdist_1(d,gh,F_k)\:=\min\{\:E\::u = semdist_1(d_1,g,F_k),$$ v =semdist_1(d_2,h,F_k), d_1d_2=d\}$.
Note that, here we take the minimum over all $d_1,d_2$ such that $d_1d_2=d$. (It is to be noted that
there may be combinations of $d_1,d_2$ for which there may not be
strings of  length $d_1$ in $L(g)$ or strings of length $d_2$ in $L(h)$; 
in these cases, it is
easy to see that  either $u\:=\infty$ or $v\:=\infty$
 respectively and hence $E\:=\infty$. Hence
these additional combinations do not change the minimum.)
Using the induction hypothesis, we have $semdist_1(d_1,g,F_k)\:=syndist(d_1,g,F_k)$
and $semdist_1(d_2,h,F_k)\:=syndist(d_2,h,F_k)$. Using this, we have
$semdist_1(d,gh,F_k)\:=\min\{\:E\::u = syndist(d_1,g,F_k),$$ v =syndist(d_2,h,F_k), d_1d_2=d\}$.
From the definitions, we see that the right hand side is $syndist(d,gh,F_k)$.
The proof of the induction step for the case when $f\:=(g)^*$ is similar and is left to the reader.
$\Box$

For a RE $f$, let $A(f)$ be a standard non-deterministic automaton 
that accepts $L(f)$ and such that the size of $A(f)$ is linear in $length(f)$ (see \cite{LP98}).
The values of $semdist_{1}(d,f,F_k)$ for each $k=1,...,\infty$ can be computed 
by constructing the automaton $A(f)$ (possibly non-deterministic) and using
the algorithm given in \cite{HS00}.
These algorithms are of complexity $O(length(d)\cdot length(f))$.

Given a database sequence $d$ and RE $f$, $semdist_{2}(d,f,F_k)$ is computed 
exactly on the same lines as given in section \ref{aut-sec}.
First we obtain the automaton $\overline{A}$ that accepts all strings in $\Delta^*-L(f)$.
The size of the resulting automaton will be $O(2^{length(f)})$.
The reminder of the steps is same as given in the section \ref{aut-sec}.
 As before, the complexity of the algorithm is triple exponential in
$length(f)$ but linear in $length(d)$.
Because of this complexity, it might be better to use
the syntactic distance measure for similarity based retrieval. Note that
this distance function is also same as the first semantic distance
function $semdist_{1}$.

\section{Related Work}
\label{rel-sec}

There have been various formalisms for representing
uncertainty (see \cite{Ha03}) such as probability measures, Dempster-Shafer belief functions,
plausibility measures, etc. 
Our similarity measures for temporal logics and automata can possibly be categorized
under plausibility measures and they are quite different from probability measures. 
The book \cite{Ha03} also describes logics for reasoning about uncertainty.
Also, probabilistic versions of Propositional Dynamic Logics were 
presented in \cite{Ko83}.
However, these works do not consider logics and formalisms on sequences, and
do not use the various vector distance measures considered in this paper.

Since the appearance of a preliminary version of this paper \cite{Si02},
other non-probabilistic quantitative versions of temporal logic have been proposed
in \cite{Al04,Al03}. Both these works consider infinite computations and 
branching time temporal logics. The similarity measure they give, for the linear time
fragment of their logic, corresponds to the infinite norm among the vector distance
functions. On the contrary, we consider formalism and logics on finite sequences
and give similarity based measures that use a spectrum vector of distance measures.
We also present  methods fo computing similarity values of a database sequence
with respect to queries given in the different formalisms.

There has been much work done on querying from time-series and other sequence
 databases. For example, methods for similarity based retrieval from 
such databases
have been proposed in \cite{FRM94,AFS93,ALSS95,B97,RM97,NRS99}. These methods 
assume that the query
is also a single sequence, not a predicate on sequences as we consider 
here. 

There has also been much work done on data-mining over time series data 
\cite{AS94,GRS99} and other databases. These works mostly consider
 discovery of patterns that have a given minimum level of support.
They do not consider similarity based
retrieval. 

A temporal query language and efficient algorithms for similarity
based retrieval have been presented in  \cite{SYV97}. That work uses a
a syntactic distance measure which is ad hoc. On the contrary, in this work, 
we consider syntactic as well as semantic distance measures. Further, 
in this paper, we consider a spectrum of these measures based on well 
accepted standard norm distance measures on vectors.

There has been work done on approximate pattern matching (see \cite{WM92}
for references)
based on regular expressions. They use different distance measures.
For example, they usually use the edit
distance as a measure and look for patterns defined by a given regular
expression with in a given edit distance. On the other hand, we
consider average measures; for example, the distance function $F_1$
defines average block distance.
In the area of bio-informatics much work has been done on
sequence matching (see \cite{D98} for references). Most of this work is based
on probabilistic models (such Markov or extended Markov models).
They do not employ techniques based on indices for the subsequence
search. 

Predicates on sequences have been employed in specifying 
triggers in Active Database
Management Systems \cite{C89,D88,GJS92,SW95a}. However, there 
exact semantics is used for firing and processing the triggers.

Lot of work on fuzzy logic considers assignment of similarity
values to propositional formulae based on their syntax. However, to the
best of our knowledge no other work has been done for logics on sequences. 
   
\section{Conclusions and Discussion}
\label{conc-sec}
In this paper, we have considered languages based on automata, temporal logic and regular
expressions for specifying queries over sequence databases. We have defined
a variety of distance measures, based on the syntax and semantics of 
the queries. We have outlined algorithms for computing these values.
The algorithms for computing syntactic distance measures are only
of polynomial time complexity in the length of the query and polynomial
in the length of the database sequence. The algorithms for computing
the first semantic distance measure have lower complexity than the
second semantic distance measure. 
Thus, from the complexity point of view, it might be better to use
the syntactic based measures or the first semantic distance measure.
Some of algorithms for automata have been implemented and tested on real data
(see \cite{HS00} for details).

It is to be noted that when we defined, in section 2, the distance $dist(d,a,F_k)$ between
a database sequence $d=(d_0,...,d_{n-1})$ and a sequence $a\:=(a_0,...,a_{n-1})$
of atomic queries taken from $\delta$, we assumed that we are given the values 
$simval(d_i,a_i)$ for 
each $i=0,..,n-1$. We also assumed that these values lie in the range ${\tt [0,1]}$.
If we require that some atomic query, say $\delta$, should be exactly satisfied 
then giving
a similarity value of either $0$ or $1$ may not achieve this purpose. Suppose
that $a_0\:=\delta$ and $d_0$ does not satisfy $a_0$; then setting $simval(d_0,\delta)=0$,
and hence setting $dist(d_0,\delta)=1$, will not serve the purpose since
$F_k$ for $k<\infty$ will aggregate these values and the value of $dist(d,a)$
may be much smaller than $1$ if other database states in the sequence satisfy
the corresponding atomic queries with similarity value $1$. (Of course, if $F_\infty$
then this is not a problem.) To achieve what we want, we need to set $simval(d_0,\delta)$
to be $-\infty$. In this case, $dist(d,a,F_k)$ will be $\infty$ for every $k>0$.
Thus for those atomic queries which need to be exactly satisfied, we can define
the similarity value of a database state with respect to these to be either $\-infty$
or $1$ denoting no satisfaction and perfect satisfaction respectively; note the
corresponding distance values will be $\infty$ or $0$ respectively.
Thus we can partition the set of atomic queries into two sets--- those that need
to be exactly satisfied for which the similarity values given are either $-\infty$
or $1$, and the remianing for which the similarity values are given from the interval
${\tt [0,1]}$. It is not difficult to see that this scheme would work for
the syntactic distance measures defined in sections 4 and 5.

It is to be  noted that all the distance measures that we defined are based on
norm vector distance functions. We feel these vector distance functions are the
most appropriate for the applications mentioned earlier in the paper.
On the other hand, other distance functions between sequences, may be 
appropriate for other applications. 
For example, the edit distance may be appropriate in applications involving
bio-informatics. As part of future work this needs further investigation.

\end{document}